\documentclass[twocolumn,journal]{IEEEtran}
\usepackage[usenames,dvipsnames]{color}
\usepackage{subfigure}
\usepackage{graphicx}
\usepackage{amsmath}
\usepackage{color}
\usepackage{multicol}
\usepackage{amssymb}
\usepackage{graphicx}

\usepackage{balance}

\usepackage{amsfonts}%
\usepackage{verbatim}%
\usepackage{enumerate}%
\usepackage{psfrag}
\usepackage{epsfig}
\usepackage{epstopdf}
\usepackage{multicol}
\usepackage{setspace}
\usepackage{dsfont}

\usepackage{algorithm}
\usepackage{algorithmic}
\usepackage{cite}
\usepackage{float}
\usepackage{cite}
 
\begin{document}
\title{Optimal Spectrum Access for a Rechargeable Cognitive Radio User Based on Energy Buffer State}
\author{\IEEEauthorblockN{Ahmed El Shafie\IEEEauthorrefmark{1}\IEEEauthorrefmark{4}\IEEEauthorrefmark{6}, Mahmoud Ashour\IEEEauthorrefmark{2}\IEEEauthorrefmark{3}\IEEEauthorrefmark{6}, Amr Mohamed\IEEEauthorrefmark{2} and  Tamer Khattab\IEEEauthorrefmark{5}}

\IEEEauthorblockA{\IEEEauthorrefmark{1} Wireless Intelligent Networks Center (WINC), Nile University, Giza, Egypt.\\
\IEEEauthorrefmark{2} Computer Science and Engineering Dept., Qatar University, Doha, Qatar.\\
\IEEEauthorrefmark{5} Electrical Engineering Dept., Qatar University, Doha, Qatar. \\
\IEEEauthorrefmark{4} Electrical Engineering Dept., University of Texas at Dallas, TX, USA. \\
\IEEEauthorrefmark{3} Electrical Engineering Dept., Pennsylvania State University, PA, USA.\\
}
\thanks{\IEEEauthorrefmark{6} Both authors contributed equally to this work.}
}
\date{}
\maketitle
\begin{abstract}
This paper investigates the maximum throughput for a rechargeable secondary user (SU) sharing the spectrum with a primary user (PU) plugged to a reliable power supply. The SU maintains a finite energy queue and harvests energy from natural resources, e.g., solar, wind and acoustic noise.
We propose a probabilistic access strategy by the SU based on the number of packets at its energy queue. We investigate the effect of the energy arrival rate, the amount of energy per energy packet, and the capacity of the energy queue on the SU throughput under fading channels. Results reveal that the proposed access strategy can enhance the performance of the SU.
\end{abstract}
\section{Introduction}

By replacing batteries of rechargeable mobile nodes, energy harvesting technology is a recently emerging promising solution to enlarge the lifetime
of wireless networks. It allows wireless devises to reuse and invest energy
from the surrounding environment \cite{c0}.

Powering a portable node is considered a very important aspect affecting the performance of wireless communication systems \cite{f1,lu2014dynamic}. In most applications, portable/mobile wireless nodes are battery-based, i.e., operate with power supplied from a battery. These batteries are limited storage capacity and frequently need to be recharged or replaced \cite{lu2014dynamic}.

Recently, several works have considered nodes with energy harvesting capability, e.g., \cite{hoang2009opportunistic,sharma2010optimal,ho2010optimal,yang2010transmission,yang2010optimal,tutuncuoglu2010optimum}. Different types of nature energy sources, such as light, vibration or heat are available to be invested by the surrounding devises \cite{survey}. The authors of \cite{hoang2009opportunistic} investigated the optimal policies for a cognitive radio devise. The problems of throughput-maximization and mean delay minimization for single-node communication are considered in \cite{sharma2010optimal}. Using a finite horizon setup, the authors of \cite{ho2010optimal} study the energy allocation for throughput-maximization.

Lately, the authors in \cite{pappas,wimob,ourletter,sultan,wcmpaper,gc2013,myletter,ElSh1502:Spectrum,workwithashour} considered a cognitive radio networks with energy harvesting capability. It is assumed that the statistics of energy arrivals follow certain random distributions. For instance, the authors of \cite{pappas,wimob,ourletter,wcmpaper,gc2013,myletter} assume that the arrived energy packets from nature are Bernoulli. However, the authors of \cite{krikidis2012stability,ElSh1502:Spectrum,workwithashour} assume Poisson arrivals.

In this work, we consider a secondary user (SU) with energy harvesting capability. We assume a simple configuration composed of one energy harvesting SU and a primary user (PU). The SU collects energy from nature with certain rate. We assume fading channels. We propose a new access strategy that depends on the exact number of packets in the energy queue. The SU probabilistically uses a certain number of energy packets based on the current available number of packets at its energy queue.  We do not assume the availability of the channel state information at the transmitters. Thus, the proposed protocol is simple and does not require the exact tracking of the channel variations. We optimize the probabilities of the used number of packets at each state to achieve the maximum secondary throughput. We compare the proposed protocol by a protocol that has been considered lately in \cite{myletter} and \cite{ElSh1502:Spectrum}. The results reveal the gains of the proposed protocol.

The rest of this paper is organized as follows. The system model under consideration is described in Section \ref{system_model}. Energy harvesting employed at the SU is studied and analyzed in Section \ref{energy_harvesting}. Section \ref{proposed_policy} presents the proposed transmission policy adopted by the SU along with its throughput analysis. Numerical results are presented in Section \ref{numerical_results}. Finally, concluding remarks are drawn in Section \ref{conclusion}.

\section{System Model}\label{system_model}
\begin{figure}[t]
\begin{center}
\includegraphics[width=1\columnwidth]{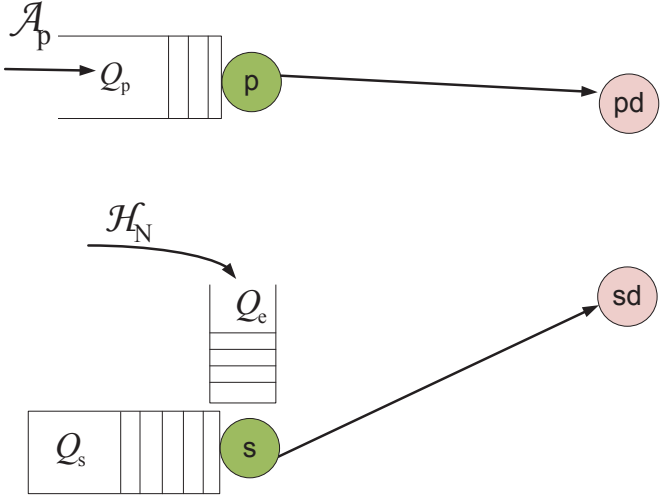}
\caption{Cognitive radio network under consideration. $\mathcal{A}_{\rm p}$ is the number of primary data packet arrivals in a given time slot and $\mathcal{H}_{\rm N}$ is the number of energy packets harvested from natural resources.}\label{fig0}
\end{center}
\vspace{-5mm}
\end{figure}

We consider a simple cognitive radio network consisted of one PU ($\rm p$) plugged to a reliable power supply and one SU ($\rm s$) equipped with energy harvesting capability. The network is shown in Fig. \ref{fig0}. The PU communicates with its destination, $\rm pd$, whenever it has packets to send, while the SU communicates with its respective destination, $\rm sd$, under certain strategy as will be explained later.
Each user maintains a data buffer to store the incoming data traffic. The data buffers are assumed to be of infinite capacity.\footnote{For a similar assumption of infinite capacity data queues, the reader is referred to \cite{pappas,wimob,ourletter,wcmpaper,gc2013,myletter,krikidis2012stability,ElSh1502:Spectrum,workwithashour} and the references therein.} The primary and the secondary data queues are denoted by $Q_{\rm p}$ and $Q_{\rm s}$, respectively. We assume fixed-length data packets of size $\beta$ bits. The SU maintains another queue, denoted by $Q_{\rm e}$, for storing the fixed-length energy packets harvested from nature. Each energy packet is assumed to contain ${\rm e}$ energy units. The capacity of $Q_{\rm e}$ is assumed to be limited to $E_{\max}$ energy packets. Thus, the energy queue contains $E_{\max} {\rm e}$ energy units at most.

Time is slotted and a slot has a duration of $T$ seconds. We do not assume the availability of the channel state information at the transmitting nodes.
The arrivals of data packets at the primary queue, $Q_{\rm p}$, are independent and identically distributed across time slots following a Bernoulli process. The mean arrival rate of $Q_{\rm p}$ is $\lambda_{\rm p}\in [0,1]$ packets/slot. The secondary data queue, $Q_{\rm s}$, is assumed to be saturated (always backlogged) with packets. The SU harvests energy from the environment (nature), e.g., solar, wind, vibration, etc. The arrivals to $Q_{\rm e}$ due to environmental energy harvesting are assumed to be distributed according to a Poisson process with rate $\lambda_{\rm e}$ energy packets/slot. For secondary data transmission, we assume that multiple energy packets may be used for a single data packet transmission. The PU has the priority to transmit if $Q_{\rm p}$ is non-empty, whereas the SU must sense the primary activity and may access the channel if the PU is inactive. The PU utilizes the whole time slot duration for its data transmission. However, the SU senses the spectrum for a duration of $\tau < T$ seconds to detect the PU's activity. The sensing process is assumed to be perfect as in \cite{krikidis2009protocol,krikidis2010stability,myletter,wcmpaper} and the references therein.

The link between every transmitter-receiver pair exhibits frequency-flat Rayleigh block fading. The channel coefficient remains constant during a time slot and changes independently from a slot to another. We denote the gain of the channel between transmitter ${\rm \nu}$ and receiver ${\rm \ell}$ ($\rm \nu\rightarrow \ell $ link) at the $t$th time slot by $h^t_{\rm \nu\ell}$.\footnote{We refer by the channel gain to the absolute squared value of the channel fading coefficient.} Since the channels are Rayleigh fading, $h^t_{\rm \nu\ell}$ is exponentially distributed with mean $\rm \sigma_{\nu\ell}$. All links are statistically independent.

The primary transmit power is assumed to be fixed and is equal to $P_{\rm p}$ Watts. An outage occurs on the link $\rm p \rightarrow pd$ when the transmission rate exceeds the link capacity. The probability that an outage occurs on that link in the $t$th time slot is given by
\begin{equation}\label{outage_p}
\mathbb{P}_{\rm ppd}=\Pr \left \{\mathcal{R}_{\rm p}> \log_2\left(1+\frac{P_{\rm p}h^t_{\rm ppd}}{\mathcal{N}_\circ W}\right)\right \}
\end{equation}
where $\Pr\{\cdot\}$ denotes the probability of the event $\{\cdot\}$, $\mathcal{R}_{\rm p}=\beta/T/W$ bits/second/Hz is the targeted primary spectral efficiency, ${\mathcal{N}_\circ}$ is the additive white Gaussian noise (AWGN) power spectral density in Watts/Hz and $W$ is the channel bandwidth in Hz. Hereafter, we omit the temporal index $t$ for simplicity.

\section{Energy harvesting}\label{energy_harvesting}
In this section, we study and analyze the arrivals of energy packets to $Q_{\rm e}$ as a result of harvesting energy from natural resources. We lay down the basis of the probabilistic energy arrivals model based on which the analysis in the rest of the paper is performed.

The energy packets harvested from nature are assumed to follow a Poisson process with rate $\lambda_{\rm e}$ \cite{krikidis2012stability}. Thus, the probability of harvesting $k$ energy packets in a given time slot is given by
\begin{equation}
\Pr\{\mathcal{H}^t_{\rm N}=k\}=\mathcal{P}_k=\frac{(\lambda_{\rm e} T)^k \exp(-\lambda_{\rm e} T)}{k!}.
\end{equation}
where $\mathcal{H}^t_{\rm N}$ is the number of energy harvested from nature in time slot $t$.

The energy queue evolves as follows:
   \begin{equation}
Q^{t+1}_{\rm e}=(Q^{t}_{\rm e}- \mathcal{D}^t_{\rm e})^+ + \mathcal{H}^{t}_{\rm N}
\end{equation}
where $(\cdot)^+$ returns the maximum between the argument and $0$, and $\mathcal{D}^t_{\rm e}$ is the number of departures from the energy queue in time slot $t$.

\section{Probabilistic Energy Transmission}\label{proposed_policy}
 The SU probabilistically selects $j$ energy packets in each time slot based on the number of packets stored in the energy queue. For instance, when the energy queue has $\mathcal{H}$ packets, the SU can use $0$, $1$, $2$, $3$, $\dots$, or $\mathcal{H}$ packets in its transmission. It also can determine the fraction of the time slots that it will use such number of packets when the energy queue is in state $\mathcal{H}$, which is equivalent to a probability of using this number of packets during that state. In other words, at each state of the energy queue, we have to assign a designable probability of using certain number of packets in data transmission that is less than or equal to the number of energy packet in the current state of the energy queue.

 The medium access control is summarized as follows:
 \begin{itemize}
 \item If the PU's queue is non-empty, the PU transmits the packet at its queue head.
 \item The SU senses the spectrum for $\tau$ seconds from the beginning of the time slot.
 \item If the PU is active, the SU remains idle till the end of the time slot.
 \item The PU's destination sends back a feedback signal to inform the PU about the decodability status of its packet.
 \item If the PU is inactive, the SU starts to transmit the packet at the head of its data queue using $0\le j\le i$ energy packets when its energy queue maintains $i\le E_{\max}$ packets. Note that the SU starts its transmission from $\tau$ seconds of the beginning of the time slot; hence, its transmission time is $T-\tau$.
     \item The SU's destination sends back a feedback signal to indicate the status of the decodability of the packet.
     \item If the intended destination cannot decode the packet, it sends back a negative-acknowledgement (NACK), and a retransmission of the packet is needed in the following time slot.
         \item If the intended receiver can decode the packet, it sends back an acknowledgement (ACK), and the packet will be removed from the transmitter's queue.
     \end{itemize}

 Let $\omega_{ij}$ denote the probability of using $j$ energy packets when the energy queue maintains $i\ge j$ energy packets.
 Since the Markov chain is \emph{aperiodic and irreducible}, it has a steady-state distribution.\footnote{\emph{irreducible means that it is possible to move from any state to any state,
and aperiodic means that it is possible to return to the same state at any
steps.}}
 Note that under the proposed protocol, the case of being silent during certain states are included when the SU uses zero energy packets for data transmission, i.e., the probability $\omega_{i0}$. This means that the SU sends nothing or equivalently remains idle.

A packet departs the primary data queue, $Q_{\rm p}$, if the channel between the PU and its respective destination is not in outage. The mean service rate of $Q_{\rm p}$ (throughput of the PU) is then given by
\begin{equation}\label{outage_p}
\mu_{\rm p}=\overline{\mathbb{P}_{\rm ppd}}=\exp\left(-\frac{\mathcal{N}_\circ W(2^{\mathcal{R}_{\rm p}}-1)}{P_{\rm p}\sigma_{\rm ppd}}\right).
\end{equation}

The probability of the PU being inactive (or the PU's queue being empty) is given by
\begin{equation}
\begin{split}
\Pi_{\rm p}\!&=\!\Pr\{Q_{\rm p}=0\}=1-\frac{\lambda_{\rm p}}{\mu_{\rm p}}
\end{split}
\end{equation}

The mean service rate of the energy queue is given by
\begin{equation}
\begin{split}
\mu_{\rm e}&=\Pi_{\rm p} \sum_{i=1}^{E_{\rm max}}  \chi_{i} \sum_{j=1}^i \omega_{ij} j
\end{split}
\end{equation}
The expression is explained as follows: The SU expends $j\le i$ energy packets with probability $\omega_{ij}$, if the PU is inactive, which occurs with probability $\Pi_{\rm p}$; and the energy queue is in state $i\le E_{\rm max}$, which occurs with probability $ \chi_{i}$.

The probability that the secondary transmission of a data packet is not in outage (successfully received at the secondary destination) is given by
\begin{equation}
\label{fgo}
\overline{\mathbb{P}_{{\rm ssd},j}}= \exp\left(-\mathcal{N}_\circ (T-\tau)\frac{2^{\mathcal{R}_{\rm s}}-1}{j {\rm e}\sigma_{\rm ssd}}\right)
\end{equation}
where $\mathcal{R}_{\rm s}=\beta/W/(T-\tau)$. Note that the successful probability of a secondary transmission is monotonically increasing with the used energy packets, $j$, and the amount of energy contained in each energy packet, ${\rm e}$.

A packet from the secondary data queue is served when the PU is inactive, and the secondary link is not in outage when the SU decides to use $j$ energy packets while its energy queue maintains $i$ packets. The throughput of the SU is given by
\begin{equation}
\begin{split}
\mu_{\rm s}&=\Pi_{\rm p}  \sum_{i=1}^{E_{\rm max}}  \chi_{i}  \sum_{j=0}^i \omega_{ij} \overline{\mathbb{P}_{{\rm ssd},j}}
\end{split}
\end{equation}
where $\Pi_{\rm p}$ represents the probability that the PU is inactive, $\chi_{i}$ represents the probability that the energy queue is in state $i$, $\omega_{ij} \overline{\mathbb{P}_{{\rm ssd},j}}$ represents the fraction of time slots, $\omega_{ij} $, that the SU uses $j$ energy packets in the transmission of its data packet and succussed in its transmission, and $\sum_{j=0}^i \omega_{ij} \overline{\mathbb{P}_{{\rm ssd},j}}$ represents the sum over all possible time fractions when the energy queue is in state $i$.

The maximum SU throughput is obtained via solving the following constrained optimization problem:
\begin{equation}
\begin{split}
& \underset{0\le \omega_{ij}\le 1 }{\max.} \,\,\,\,\,\,\ \mu_{\rm s}=\Pi_{\rm p}  \sum_{i=1}^{E_{\rm max}}  \chi_{i}  \sum_{j=0}^i \omega_{ij} \overline{\mathbb{P}_{{\rm ssd},j}} \\& \,\,\,\,\,\ {\rm s.t.} \ \ \sum_{j=0}^{i} \omega_{ij}=1~ \forall i \in \{0,1, \hdots, E_{\max}\}.
\end{split}
\end{equation}

It what follows, we compute the steady state distribution of the energy queue, i.e., $\{\chi_i\}_{i=0}^{\infty}$, to completely characterize the SU throughput for a given $\omega_{ij}$. Towards this objective, we model the evolution of the energy queue with a Markov chain. The probability of transition from state $n$ to state $k$, denoted by $P_{n \rightarrow k}$, is given by
\begin{enumerate}
\item For $k < E_{\max}$,
\begin{itemize}
\item at $k < n$,
\begin{equation}
P_{n \rightarrow k}=\Pi \left[ \displaystyle \sum_{m=n-k}^{n} \omega_{nm}\mathcal{P}_{k-(n-m)} \right]
\end{equation}
\item at $k \geq n$,
\begin{equation}
P_{n \rightarrow k}=\Pi \left[ \displaystyle \sum_{m=0}^{n} \omega_{nm}\mathcal{P}_{k-(n-m)} \right]
+
\overline{\Pi}\mathcal{P}_{k-n}.
\end{equation}
\end{itemize}
\item For $k=E_{\max}$,
\begin{align}
P_{n \rightarrow k}&=\Pi \left[\displaystyle \sum_{m=0}^{n} \omega_{nm}
\left(1-\displaystyle \sum_{\ell=0}^{k-(n-m)-1} \mathcal{P}_{\ell} \right) \right] \notag \\
&+
\overline{\Pi} \left[1-\displaystyle \sum_{\ell=0}^{k-n-1} \mathcal{P}_{\ell} \right].
\end{align}
\end{enumerate}
The $(E_{\max}+1) \times (E_{\max}+1)$ transition probability matrix, denoted by $\Lambda$, is constructed according to the above description. The $1 \times E_{\max}$ steady state distribution vector $\boldsymbol{\chi}=[\chi_0,\chi_1,\dots,\chi_{E_{\max}}]$ is obtained through solving
\begin{equation}
\boldsymbol{\chi}=\boldsymbol{\chi} \Lambda.
\end{equation}

\begin{figure}[t]
\begin{center}
\includegraphics[width=1\columnwidth , height=0.7\columnwidth]{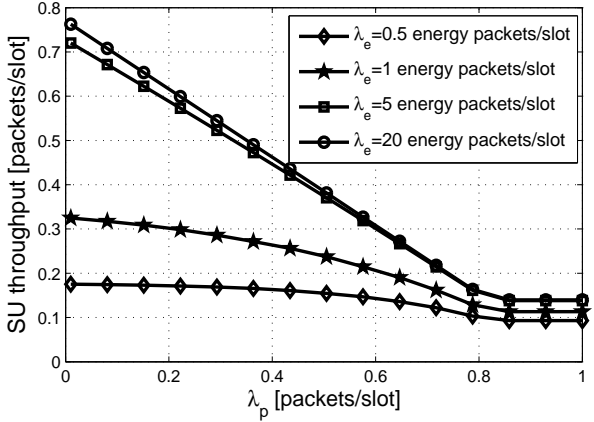}
\caption{SU throughput versus $\lambda_{\rm p}$ for different values of $\lambda_{\rm e}$.} \label{fig1}
\end{center}
\vspace{-5mm}
\end{figure}

\section{Numerical Results}\label{numerical_results}
In this section, the maximum secondary throughput is evaluated under the proposed probabilistic transmission strategy. We investigate the impact of some important factors on the achievable throughput for the SU. That is,
\begin{enumerate}
\item We study the impact of the bursty arrivals at the PU's queue on the SU throughput.
\item We investigate the effect of harvesting energy from natural resources on the secondary throughput.
\item The roles of the maximum capacity of the energy queue, $E_{\max}$, the amount of energy per energy packet, ${\rm e}$, and the energy arrival rate, $\lambda_{\rm e}$, are all highlighted in this section.
\end{enumerate}

The results are drawn using the following common parameters: The data packets have fixed length of $\beta=10^3$ bits. Time slot duration $T$ is set to $1$ second, while the sensing duration $\tau=0.1$ second. We note that $10 \%$ of the time slot is used by the SU to perform sensing which to some extent justifies the assumption of perfect sensing.
Channel bandwidth $W=1$ KHz. The noise power spectral density $\mathcal{N}_{\circ}=10^{-6}$ Watts/Hz. An energy packet contains $10^{-3}$ joules, i.e., ${\rm e}=10^{-3}$ joules, unless otherwise stated. The variances of both links $\rm p \rightarrow s$ and $\rm s \rightarrow sd$ are set to unity, i.e., $\rm \sigma_{ps}=\sigma_{ssd}=1$. However, we consider $\rm \sigma_{ppd}=0.5$.

\begin{figure}[t]
\begin{center}
\includegraphics[width=1\columnwidth , height=0.7\columnwidth]{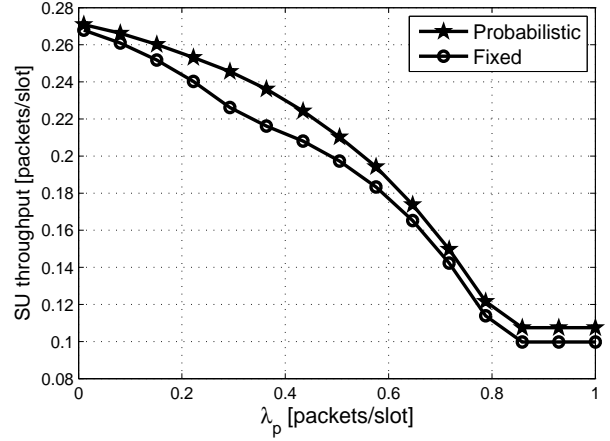}
\caption{SU throughput versus $\lambda_{\rm p}$ for fixed and probabilistic strategies.} \label{fig2}
\end{center}
\vspace{-5mm}
\end{figure}

In Fig. \ref{fig1}, we investigate the dependence of the SU throughput on energy packets' arrival rate at $Q_{\rm e}$, $\lambda_{\rm e}$. We note that the SU throughput is enhanced as $\lambda_{\rm e}$ increases. This is attributed to the increase in the availability of energy at the SU required to support its packets transmission. That is, this increases the possibility of having high number of energy packets when the PU is inactive and therefore boosts the possibility of successful decoding of the data packet at the secondary destination. The figure is plotted using the common parameters, $E_{\max}=4$ energy packets and the values of $\lambda_{\rm e}$ in the figure's legend. In Fig. \ref{fig2}, we demonstrate the advantage of the proposed probabilistic strategy over the fixed access strategy investigated in \cite{myletter} and \cite{ElSh1502:Spectrum}. In the fixed strategy scheme, the SU always uses $\mathcal{G}$ energy packets for the transmission of its data packet when that number of energy packets are available at the energy queue. The figure is generated using the common parameters, $E_{\max}=3$ energy packets, and $\lambda_{\rm e}=0.5$ energy packets/slot.

\begin{figure}[t!]
\begin{center}
\includegraphics[width=1\columnwidth , height=0.7\columnwidth]{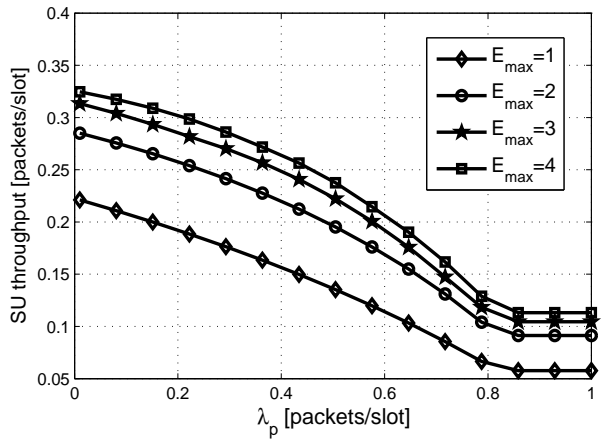}
\caption{SU throughput versus $ \lambda_{\rm p}$ for different values of $E_{\max}$.}\label{fig3}
\end{center}
\vspace{-5mm}
\end{figure}

\begin{figure}[t!]
\begin{center}
\includegraphics[width=1\columnwidth , height=0.7\columnwidth]{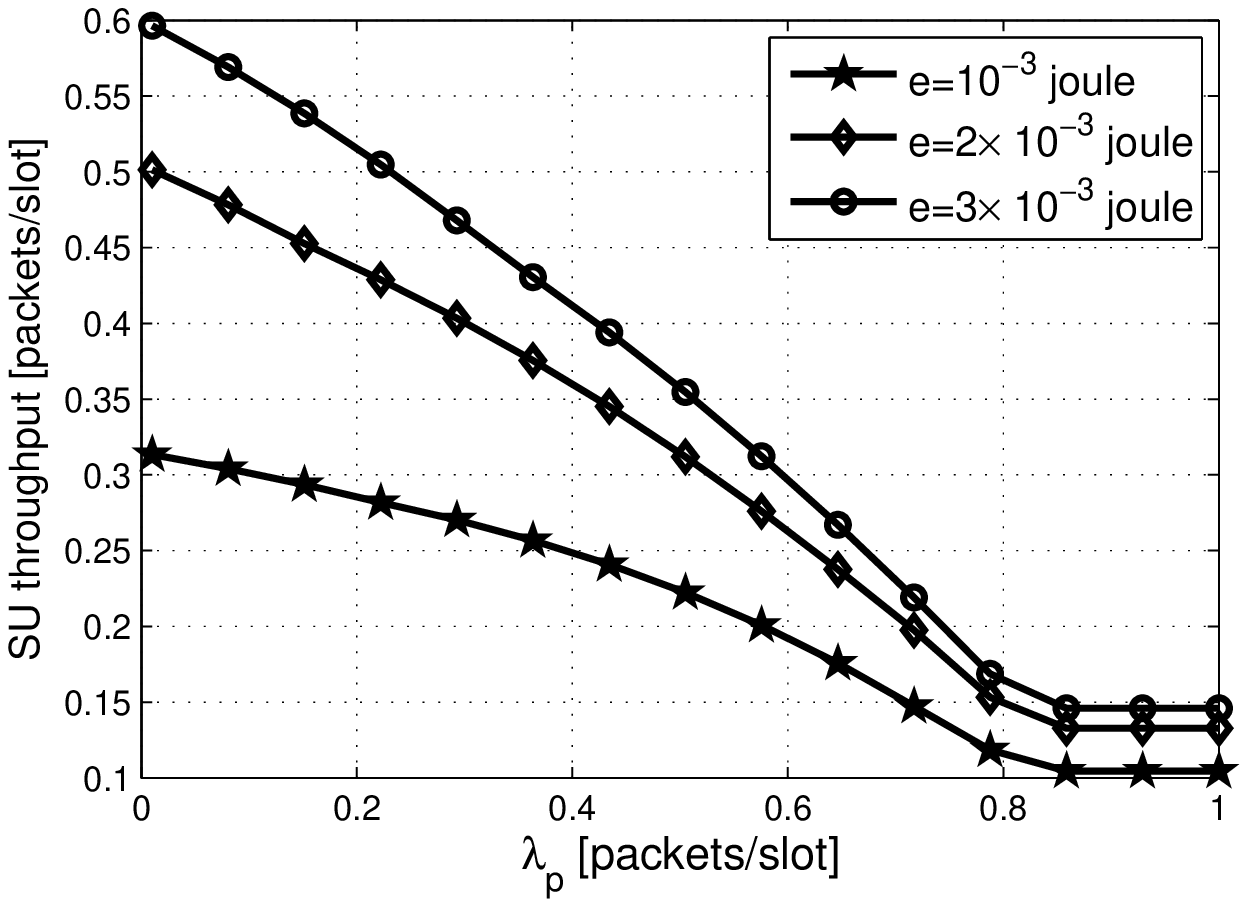}
\caption{SU throughput versus $\lambda_{\rm p}$ for different values of ${\rm e}$.}\label{fig4}
\end{center}
\vspace{-5mm}
\end{figure}

In Fig. \ref{fig3}, we show the impact of the energy buffer size on the secondary achievable throughput. As the capacity of the buffer increases, the secondary throughput increases as well. This is because the SU will be able to maintain more packets that can be used in future transmissions. The figure is drawn using the common parameters, $\lambda_{\rm e}=1$ energy packets/slot, and $E_{\max}$ in the figure's legend. Fig. \ref{fig4} demonstrates the impact of ${\rm e}$, which represents the amount of energy contained in an energy packet, on the secondary throughput. As expected, when ${\rm e}$ increases, the throughput increases as well. This is because as ${\rm e}$ increases, the number of collected packets for a single data transmission will decrease relative to the case where ${\rm e}$ has a lower value. Hence, we can achieve higher successful probability for the secondary transmission (the outage probability decreases with ${\rm e}$ as mentioned beneath (\ref{fgo})) with lower number of packets in the energy queue; hence, the SU will not wait for long time to achieve the same performance for lower ${\rm e}$. The figure is generated using the common parameters, $E_{\max}=3$ energy packets, and $\lambda_{\rm e}=1$ energy packets/slot.

\section{Conclusion}\label{conclusion}

In this work, we have investigated the maximum achievable throughput for a rechargeable SU. The SU has a limited-capacity battery and is equipped with energy harvesting capability. We assume that the SU harvests energy from natural resources. We have proposed a probabilistic access strategy based on the number of packets in the secondary energy queue. We have investigated the effect of the energy arrival rate, the amount of energy per energy packet, and the capacity of the energy queue on the SU throughput.

A possible extension of this work is to consider nodes with multiple antennas. Moreover, we can consider channel state information at the transmitters and study its impact on the performance of the system. We can also assume cooperation between the PU and the SU under the proposed protocol.
\balance

\bibliographystyle{IEEEtran}
\bibliography{IEEEabrv,term_bib}
\balance

\end{document}